\documentclass[a4paper,11pt]{article}
\usepackage[utf8]{inputenc}
\usepackage[T1]{fontenc}
\usepackage{graphicx,url}
\usepackage{a4wide}
\usepackage{mathpazo}
\usepackage[font=footnotesize,labelfont=bf]{caption}
\usepackage{amsmath}
\usepackage{bm}
\usepackage{amssymb}
\usepackage{multirow}
\usepackage{color,soul}
\usepackage{xurl}
\usepackage{authblk}
\usepackage[hidelinks]{hyperref}
\usepackage{adjustbox}
\usepackage[shortlabels]{enumitem}
\setlist[enumerate]{nosep}
\usepackage[table,xcdraw]{xcolor}
\usepackage{tabulary}
\usepackage{booktabs}
\usepackage[numbers,compress]{natbib}

\usepackage{float} 

\usepackage{verbatim}
\newcommand{\detailtexcount}[1]{%
  \immediate\write18{texcount -merge -sum -q main.tex > main.wcdetail }%
  \verbatiminput{main.wcdetail}%
}

\definecolor{Magenta}{rgb}{0.8, 0.1, 0.6}

\definecolor{LightBlue}{rgb}{0.25, 0.55, 0.9}

\title{The limits of visitation entropy as a summary\\of mobility patterns}

\author[1]{Sílvia de Sojo\thanks{Corresponding author: \href{mailto:sdsc@dtu.dk}{sdsc@dtu.dk}}}
\author[1,2]{Sune Lehmann}
\author[1]{Laura Alessandretti}
\affil[1]{DTU Compute, Technical University of Denmark}
\affil[2]{Center for Social Data Science, University of Copenhagen}

\begin{document}
\maketitle
\begin{abstract}
    Visitation entropy, the Shannon entropy of an individual’s distribution of visits across locations, is a widely used metric in the human mobility literature.
    Yet, its widespread use rests on assumptions that are rarely made explicit: entropy is defined over a fixed set of states, and estimating it empirically requires abundant, well-sampled observations.
    The limitations that arise when entropy is used outside this setting have been documented and explored in fields such as statistical physics, ecology, and cryptography.
    The implications for mobility studies, however, remain unclear.
    Here, we leverage synthetic and empirical trajectories to systematically examine the strengths and weaknesses of visitation entropy as a measure for characterizing human mobility.
    We show that, for a sequence of locations visited by an individual, the visitation entropy primarily reflects the number of unique locations visited and sequence length, which together explain 90.7\% of its variance in empirical data.
    We also show that shorter sequences systematically lead to an underestimation of entropy, showing finite-sample bias to be a key limitation.
    As a result, comparisons of groups based on visitation entropy should be treated with caution.
    We find that apparent entropy gaps between genders, commuters and non-commuters, and urban and rural residents are reduced, or reversed, once sequence length and the number of unique locations are taken into account.
    Finally, we show that these issues can be addressed by controlling for sequence length and the number of unique locations, and by complementing entropy with structural network measures, which provide more nuanced insight into how mobility is organized beyond the aspects that entropy alone captures.

\end{abstract}


\section*{Introduction}
Over the past two decades, individual trajectory data collected from mobile phones has transformed human mobility research, enabling population-scale observation of movement with unprecedented detail~\cite{barbosa-2018-human}.
Individual trajectories capture rich information about where people go, how often, and how places connect.
A central challenge is therefore to distill this high-dimensional information into summary measures that are simple enough to allow comparison within and across individuals, yet rich enough to preserve meaningful differences.

Among the indicators developed in the field, few have been as widely adopted as the \emph{visitation entropy}, also known as uncorrelated entropy or mobility entropy~\cite{song-2010-limits}.
Visitation entropy quantifies the expected information needed to determine an individual’s location from the distribution of their past visits~\cite{song-2010-limits}.
It was introduced as an adaptation of Shannon entropy~\cite{shannon-1948-mathematical}, which in information theory quantifies the uncertainty of a probability distribution over discrete outcomes, such as symbols produced by a communication source.
Its simplicity has made the visitation entropy a standard tool to measure the diversity of mobility patterns across applications, from individual-level analyses of exploration and behavioral change~\cite{poudyal-2024-dynamic, wang-2020-entropybased, vanni-2024-aggregated} to population-level comparisons across sociodemographic groups, urban environments, and socio-economic strata~\cite{gauvin-2020-gender, reisch-2021-behavioral, bertocchi-2025-big, jiang-2024-entropybased, yuan-2016-analyzing, pappalardo-2016-analytical, vanni-2024-aggregated}.

Yet this widespread use rests on assumptions that are rarely made explicit.
In communication systems, entropy is defined over a fixed alphabet and is typically estimated from abundant, well-sampled observations.
Applying it beyond that scope introduces well-documented limitations across domains such as statistical physics, ecology, machine learning, and cryptography~\cite{schurmann-2004-bias, grassberger-2008-entropy, quinlan-1996-improved, haegeman-2008-limitations, morrison-2023-mathematical, skorski-2015-shannon, koc-2023-development, gotelli-2001-quantifying, roswell-2021-conceptual}.
A key problem is that when Shannon entropy is estimated from small samples, rare events are not captured, making systems appear more regular than they are~\cite{schurmann-2004-bias, grassberger-2008-entropy, schurmann-2015-note, osgood-2016-theoretical}.
Mobility data are particularly vulnerable: individual trajectories are typically short, spanning weeks or months, yielding few observations --- a recent study based on GPS resolution data reports that a month-long trajectory has an average of 80 visits across 20 locations~\cite{desojo-2026-womens} --- a scarcity that makes entropy estimates prone to small-sample bias.
The effective number of states can also vary with spatial resolution, so that fine-grained data capture many distinct places while coarser data blur them together~\cite{vanhoof-2018-comparing}.
Shannon entropy is also challenging to interpret because it characterizes a trade-off between two distinct factors: how many states are possible and how likely each state is.
Whether and how these features shape visitation entropy estimates in mobility research remains an open question.

In this work, we characterize the visitation entropy using two data sources.
Using synthetic trajectories generated with the Exploration and Preferential Return model~\cite{song-2010-modelling}, we disentangle the effects of repertoire size (number of locations), activity (sequence length), and visit heterogeneity (the concentration of visits across locations) on entropy estimates under controlled conditions.
We then analyze a large-scale GPS dataset from 543,155 smartphone users across ten countries~\cite{alessandretti-2020-scales, desojo-2026-womens}, and test whether entropy reflects structural features of mobility by representing trajectories as networks of locations connected by sequential visits.
Finally, we examine the implications of these limitations for sociodemographic comparisons --- by gender, commuting status, and urbanization level --- revealing how analyses based on entropy alone can distort conclusions about group differences.

\section*{Results}

Shannon entropy is a property of a discrete probability distribution $p(i)$, quantifying the average information required to identify an outcome from the set of possible states $i$~\cite{shannon-1948-mathematical}.
In practice, it is estimated from empirically observed event frequencies $f(i)$, an approximation of the underlying data-generating distribution~\cite{schurmann-2004-bias, grassberger-2008-entropy, schurmann-2015-note}.

In human mobility, data are often represented as sequences of visits to distinct locations.
For an individual, the \emph{visitation entropy} corresponding to a given sequence is defined as
\[
    \hat{S} \;=\; -\sum_{i=1}^{\hat{k}} f(i)\,\log_2 f(i),
\]
where $f(i)$ is the fraction of visits to location $i$ and $\hat{k}$ is the number of unique locations visited (the location repertoire size).
Thus, the visitation entropy, $\hat{S}$, is the Shannon entropy of the empirical visit distribution $f(i)$~\cite{shannon-1948-mathematical}.
For simplicity, we will refer to visitation entropy simply as \emph{entropy} in the following.

In information-theoretic terms, $\hat{S}$ is the average number of binary questions (bits) needed to determine a user’s current location from their visit history.
When all locations are visited equally often, uncertainty is maximal.
For example, a user visiting eight locations equally requires on average $\hat{S} = -8 \cdot \tfrac{1}{8}\log_2\tfrac{1}{8} = 3$ bits.
In this case, each bit of information is used to systematically divide the set of locations into two equal parts, ensuring that each choice narrows the possibilities until only a single location (state) is left.
When visits are not uniformly distributed, a targeted questioning strategy --- starting with the most frequent location --- reduces the average number of questions, yielding lower entropy.
Formally, the entropy is the minimum possible average number of yes/no questions needed to identify the relevant state for any distribution.

Larger values of $\hat{S}$ therefore indicate more uniformly distributed visits (greater irregularity, lower predictability), while lower values reflect concentration in a few locations (greater regularity, higher predictability). Entropy is therefore interpreted as a measure of mobility regularity, providing a concise descriptor of how broadly and uniformly individuals distribute their visits across locations.

Yet, research in other fields shows that applying Shannon entropy comes with important caveats~\cite{schurmann-2004-bias, grassberger-2008-entropy, quinlan-1996-improved, haegeman-2008-limitations, morrison-2023-mathematical, skorski-2015-shannon, koc-2023-development}. How these caveats impact mobility analyses has heretofore been largely unexplored.

In the following, we examine three aspects of visitation entropy that challenge its interpretation in mobility science using synthetic~\cite{song-2010-modelling} and empirical mobility sequences (Methods:~\nameref{sec:meth-data}).

\subsection*{Limitation 1: The intertwined role of repertoire size and visit heterogeneity}

By definition, the \emph{visitation entropy} $\hat{S}$ depends on both the \textit{heterogeneity} in the visit distribution and the size of the \textit{repertoire} of locations.
This interdependency carries a key implication: interpreting entropy alone is challenging, as the sequences corresponding to two individuals may result in the same entropy while arising from very different mobility behaviors.

Let's take a minimal example: consider two individuals, individual A visits 10 locations with a highly skewed distribution (e.g., 56\% of visits to one place, sampled from a power-law with exponent $1.5$), while individual B visits only 5 locations with a more flat frequency distribution (power-law with exponent $0.6$, see Fig.~\ref{fig:lim12}a left).
Despite differing repertoire sizes and visit distributions, both yield identical entropy values ($\hat{S}\approx 2.3$ bits).

To generalize this observation, we simulate synthetic mobility sequences using an established generative model~\cite{song-2010-modelling}. Following the literature, at equilibrium, a mobility trajectory can be generated using the preferential return mechanism (see Methods: \nameref{sec:methods-epr}).

In this framework, an individual’s mobility is defined by three key parameters: the maximum size of their \emph{repertoire of locations} ($k$), the \emph{heterogeneity} in visit distribution across locations ($\alpha$), and their \emph{activity} level --- or sequence length --- ($N$).
Here, heterogeneity captures how flat or peaked the visit distribution is across locations, with higher $\alpha$ corresponding to more peaked (more heterogeneous) distributions.
The probability of visiting a location $i$ follows a power-law distribution $p(i)\sim i^{-\alpha}$, characterized by parameter $\alpha$.
We generate sequences of $N$ visits to $\hat{k}\leq k$ distinct locations by sampling visits from the power-law distribution.
The entropy of the resulting visit distributions can then be computed either empirically, $\hat{S}$, from the observed visit frequencies $f(i)$, or analytically, $S$, from the underlying distribution $p(i)$ (see Methods for model details and Supplementary Material~\ref{si:numval} for validation of the simulation framework).

We first study how the analytical entropy $S$ depends on $k$ and $\alpha$.
$S$ decreases with a more peaked visit distribution (increasing $\alpha$) and with smaller repertoire sizes ($k$), reflecting their intertwined effects (see Fig.~\ref{fig:lim12}b top).
Additionally, the entropy is more sensitive to changes in $\alpha$ as $k$ increases, highlighting that entropy is not scale invariant (see Supplementary Material~\ref{si:Ssens}).
This has the important consequence that differences in entropy cannot be interpreted in the same way for individuals visiting a different number of unique locations.

The role of repertoire size has been partially recognized in prior studies \cite{bertocchi-2025-big, macedo-2022-differences, centellegher-2025-job, vanhoof-2018-comparing}.
To isolate the effects of the heterogeneity in the visit distribution, one approach has been to normalize entropy by its maximum value, $\log(k)$~\cite{cover-1991-entropy}, leading to the normalized form $S/\log(k)$~\cite{macedo-2022-differences, centellegher-2025-job, montjoye-2016-bandicoot}.
Other, less common normalizations include normalizing by $\log(N)$ \cite{pappalardo-2015-using, montjoye-2016-bandicoot}, adding  $\frac{k-1}{2N}$~\cite{degregorio-2024-entropy, paninski-2003-estimation} or applying cell-tower-specific corrections in call detail records~\cite{vanhoof-2018-comparing}.
However, as we demonstrate in Figure~\ref{fig:lim12}b bottom, and in Supplementary Material~\ref{si:Ssens}, current normalization approaches, including normalizing by $\log(k)$, do not fully remove the dependence of estimated entropy $S$ on repertoire size $k$.

\subsection*{Limitation 2: The dependence on the sequence length} 
\begin{figure}[H]
    \centering
    \includegraphics[width=1\linewidth]{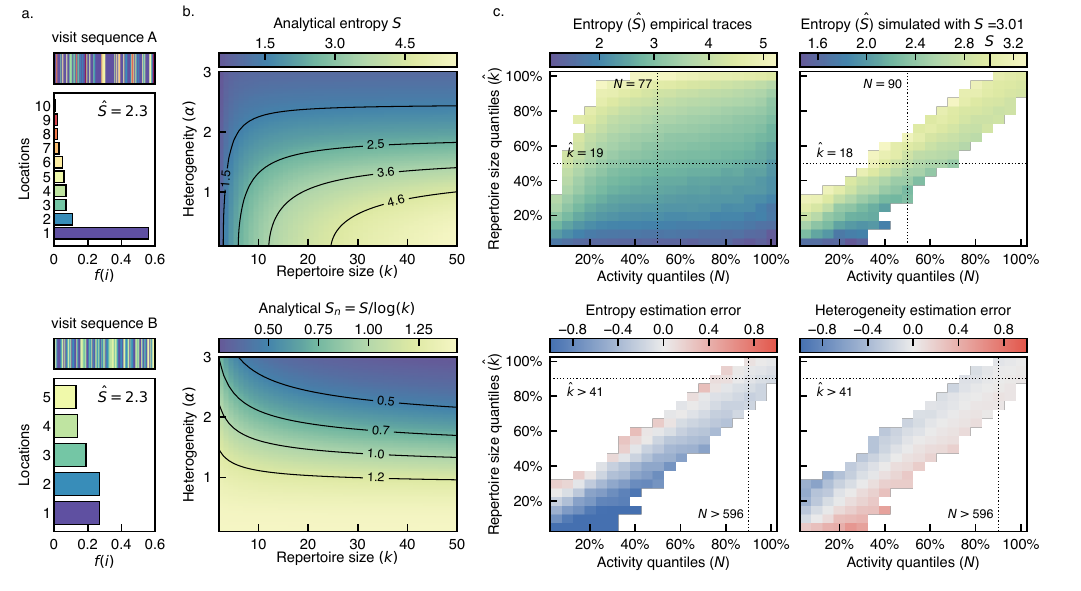}
    \caption{
        \textbf{Visitation entropy conflates the effects of repertoire, visit heterogeneity and sequence length: a. }
        Two example mobility traces (A and B) with identical entropy ($\hat{S} = 2.3$) but drawn from different visit distributions and covering different total numbers of locations.
        \textbf{b.} Top: Analytical entropy ($S$, color scale) as a function of heterogeneity ($\alpha$) and repertoire size ($k$), showing that the same entropy value can emerge from distinct behavioral patterns. Bottom: Entropy normalized by $log_2(k)$. Normalization reduces but does not eliminate dependence on repertoire size, particularly at small $k$.
        \textbf{c. }
        Four heatmaps showing the relationship between activity ($N$), repertoire size ($\hat{k}$), and visitation entropy ($\hat{S}$) in empirical and synthetic mobility traces.
        Top-left: Entropy ($\hat{S}$) in empirical traces across quantiles of $N$ and $\hat{k}$. Dotted lines indicate median empirical values.
        Top-right: Entropy from synthetic traces generated with fixed analytical entropy ($S = 3.01$), heterogeneity ($\alpha = 1.65$), and repertoire size ($k = 60$), across increasing $N$ (see other configurations in Supplementary Material~\ref{si:lim2-otherconfigs}).
        Bottom-left: Entropy estimation error ($\hat{S} - S$) in synthetic traces.
        Bottom-right: Heterogeneity estimation error ($\hat{\alpha} - \alpha$) in synthetic traces.
    }
    \label{fig:lim12}
\end{figure}

Entropy is a property of the visit distribution and, theoretically, it is independent of the sequence length $N$.
In practice, however, one needs enough data to estimate the entropy accurately.
Indeed, our empirical data (see Methods: \nameref{sec:meth-data}) reveal a correlation between $\hat{S}$ and $N$, even when the number of unique locations $\hat{k}$ is held constant (see Fig.~\ref{fig:lim12}c, top-left).
Specifically, for a fixed $\hat{k}$, longer traces are associated with lower entropy values.

To investigate this dependency, we turn to synthetic data.
We simulate sequences of varying length $N$ using a generative model with fixed parameters $\alpha = 1.65$ and $k = 60$, resulting in $S=3.01$ (Fig.~\ref{fig:lim12}c, top-right; see results for other configurations in Supplementary Material~\ref{si:lim2-otherconfigs}).

Although all sequences are drawn from the same underlying distribution and share the same analytical entropy $S$, empirical estimates of $\hat{S}$ systematically underestimate $S$, converging only for large $N$ (above the 90th percentile, $N > 596$; see Fig.~\ref{fig:lim12}c bottom-left).
For a fixed repertoire size $\hat{k}$, as $N$ grows the observed frequency distributions become increasingly right-skewed; when $\hat{k}$ is large, they more closely reflect the underlying visit heterogeneity $\alpha$ of the model (Fig.~\ref{fig:lim12}c, bottom-right; see Methods for details on estimating $\hat{\alpha}$).
This bias arises from the finite-sample effects common to heavy-tailed distributions, where short sequences rarely capture low-probability locations, leading to statistical fluctuations in the distribution of visits and biased entropy estimates~\cite{schurmann-2004-bias, grassberger-2008-entropy}.

In practice, mobility studies often rely on observation windows ranging from a few weeks to several months.
In our dataset, for example, trajectories span one month, with a median sequence length of $N = 77$.
Given that the median repertoire size is $\hat{k} = 19$, the number of states is comparable to the number of observations --- precisely the regime where entropy underestimation is most pronounced~\cite{schurmann-2015-note}.
Importantly, this bias is not resolved by normalization by $\log(\hat{k})$ or by the Miller-Madow correction $\frac{k-1}{2N}$, despite the latter explicitly accounting for sequence length (see Supplementary Material~\ref{si:lim2-othernorm}).
Together, our results show that controlling for $\hat{k}$ alone is insufficient to isolate the impact of $N$ on entropy estimates.

To quantify the extent to which $\hat{S}$ reflects simple statistics, we trained a random forest regression model (details in Methods:~\nameref{sec:meth-regression}) to explain entropy using only two predictors: the number of unique locations visited (repertoire size, $\hat{k}$) and the total number of visits (activity level, $N$).
This model explains $90.7\%\pm0.27$ of the variance in entropy across individuals in our empirical data (see Supplementary Material~\ref{si:lim3-xgboost}), revealing that the metric is largely driven by trivial differences in activity and location count.
This implies that direct comparisons of entropy across individuals with different $N$ and $\hat{k}$ can be misleading.

\subsection*{Limitation 3: The entropy does not capture structural properties of mobility networks}
Even when activity and repertoire size are controlled for, entropy does not account for the structural organization of movement, such as how locations are connected through transitions and how visits are sequenced over time.
To examine such structures, we represent mobility sequences as \emph{individual mobility networks}: directed weighted networks where nodes correspond to locations, edges link consecutive visits, and edge weights capture the number of transitions (see example networks in Fig.~\ref{fig:lim3}a).
When comparing sequences of equal length (e.g., one month), this framework enables us to examine structural differences in mobility using network metrics (Methods:~\nameref{sec:methods-netw},~\nameref{sec:meth-nwmet})~\cite{desojo-2026-womens}.
Here, we apply the network framework to assess how structural movement properties connect to the entropy $\hat{S}$.

We find that some structural aspects of mobility networks correlate with $\hat{S}$ because they are inherently tied to the heterogeneity of the visit distribution.
For example, entropy is negatively correlated with the degree centrality of the most visited location (with partial correlation controlling for $\hat{k}$: $\rho_{\#1deg,\hat{S}\cdot \hat{k}} = -0.383$, p-value $p < 10^{-16}$; see Methods for details), reflecting the concentration of visits at a dominant hub location.
Yet entropy, which is only designed to present a trade-off between number of states and how an individual distributes visits among them, is unable to capture important structural features, and shows little association with density ($\rho_{\text{density},\hat{S} \cdot \hat{k}} = 0.058$, $p < 10^{-16}$), clustering coefficient ($\rho_{\text{Clust},\hat{S} \cdot \hat{k}} = -0.112$, $p < 10^{-16}$) and the number of cycles ($\rho_{\text{Cycles},\hat{S} \cdot \hat{k}} = 0.058$, $p < 10^{-16}$).
This insensitivity is clear in Fig.~\ref{fig:lim3}b (top-right), where entropy remains stable across wide variation in clustering and cycles. Fig.~\ref{fig:lim3}a further illustrates this limitation: networks with nearly identical entropy, activity, and repertoire size can be weakly clustered with few cycles (blue) or highly clustered and cycle-rich (red).

Including network features alongside activity and repertoire size in a regression model raises the explained variance from $90.73\%\pm0.27$ to $95.26\%\pm0.09$, with the centrality of the top-degree node as the dominant contributor (see Supplementary Material~\ref{si:lim3-xgboost}).
Entropy thus reflects visit concentration but not the structural complexity that distinguishes individual mobility patterns.

\subsection*{Temporal structure alone cannot overcome visitation entropy’s limitations}
A potentially better candidate for capturing more structural information in the visit sequences is the temporal entropy, $\hat{S}^{temp}$ (see Methods and \cite{song-2010-limits} for the definition), which accounts for the probability of observing specific time-ordered subsequences $T_i'$ within a trajectory \cite{song-2010-limits}.
In principle, $\hat{S}^{temp}$ should precisely reflect higher-order regularities in mobility patterns.

\begin{figure}[H]
    \centering
    \includegraphics[width=1\linewidth]{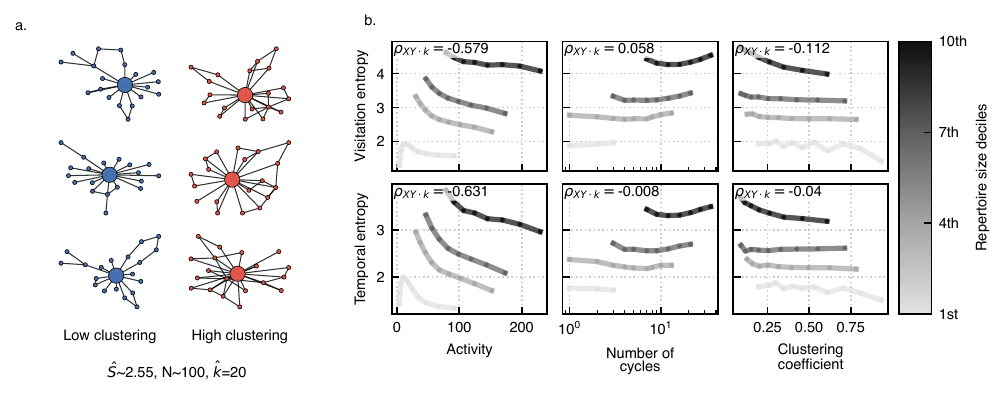}
    \caption{\textbf{Entropy does not capture structural differences in mobility. a.}
        Example mobility networks with similar visitation entropy ($\hat{S}=2.55\pm0.02$), activity ($N=100\pm2$), and repertoire size ($\hat{k}=20$), but contrasting clustering and number of cycles: low (blue) versus high (red). Nodes represent locations, sized by visit frequency (scaled exponentially for visibility), and positioned by geographical distance; links denote sequential visits. Networks were chosen to highlight the patterns in (b): low clustering with few cycles (blue) versus high clustering with many cycles (red). \textbf{b.} Empirical visitation entropy ($\hat{S}$, top) and temporal entropy ($\hat{S}^{temp}$, bottom) across activity, number of cycles, and clustering, with repertoire size ($\hat{k}$) held fixed (quantiles shown in grey). Similar entropy values can emerge under very different structural conditions. Partial correlations controlling for repertoire size are reported in the upper left of each panel.
    }
    \label{fig:lim3}
\end{figure}

However, we find that $\hat{S}^{temp}$ faces limitations similar to visitation entropy (see Fig.~\ref{fig:lim3}b bottom).
It remains strongly influenced by repertoire size (correlation $\rho_{k,\hat{S}^{temp}} = 0.759$, $p < 10^{-16}$) and activity level ($\rho_{N,\hat{S}^{temp} \cdot k} = -0.631$, $p < 10^{-16}$; see Fig.~\ref{fig:lim3}c), and it shows a negligible correlation with structural features such as the number of cycles ($\rho_{\text{Cycles}, \hat{S}^{temp} \cdot \hat{k}} = 0.008$, $p < 10^{-3}$) and the clustering coefficient ($\rho_{\text{Clust}, \hat{S}^{temp} \cdot \hat{k}} = 0.040$, $p < 10^{-16}$).

These limitations stem from $\hat{S}^{temp}$ relying on compression-based estimation requiring long and well-sampled trajectories to converge~\cite{schurmann-2015-note, grassberger-2008-entropy}.
As shown in Limitation 2, empirical trajectories are too short to recover the visit frequency distribution, and are likewise inadequate to capture the structural richness that temporal entropy promises.

\subsection*{Implications for mobility research}
The limitations identified above have direct consequences for empirical mobility research.
When populations differ in activity levels or repertoire sizes ($N$, $\hat{k}$), entropy-based comparisons risk distorting conclusions about group differences.
We illustrate this in detail for gender differences before examining whether similar patterns arise when comparing commuters with non-commuters or urban with rural populations.

\textbf{Gender differences.}
Previous research has shown that women tend to make fewer visits than men but typically explore more distinct locations~\cite{desojo-2026-womens}.
Here, we characterize the average difference between male and female activity, repertoire size, and entropy using the relative difference with respect to their mean ($\frac{X_{\text{Male}} - X_{\text{Female}}}{(X_{\text{Male}} + X_{\text{Female}})/2}$), estimated via bootstrapping.

\begin{figure}[H]
    \includegraphics[width=0.5\linewidth]{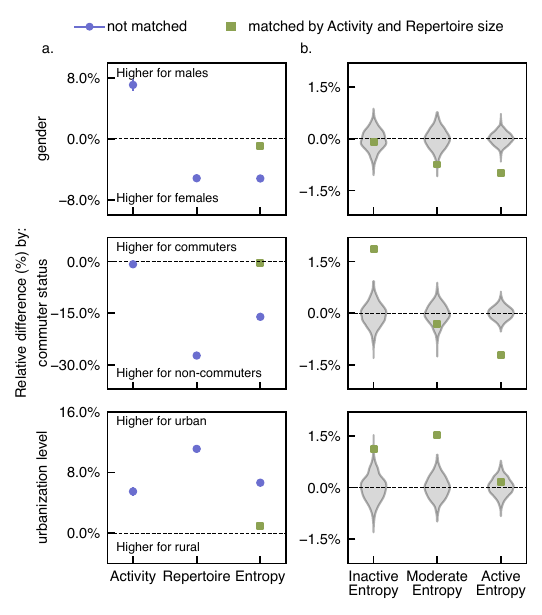}
    \caption{\textbf{The implications of entropy’s limitations for sociodemographic comparisons.}
        \textbf{a.} Relative differences in median activity (N), repertoire size ($\hat{k}$), and visitation entropy ($\hat{S}$) between groups. Blue points show unmatched comparisons, estimated as the mean relative difference in medians with bootstrapped standard errors.  Green squares show results after matching individuals by activity and repertoire size (see Methods).
        Rows correspond to comparisons by gender (top), commuter status (middle), and urbanisation level (bottom). \textbf{b.} Same as a, but differences are stratified by activity level (inactive, moderate, active). Grey violin plots represent reference distributions from 1,000 random permutations of sociodemographic labels.
    }\label{fig:sn-demo}
\end{figure}

In line with prior work, we find that women have lower median activity levels but higher median repertoire sizes, resulting in higher median visitation entropy (Fig.~\ref{fig:sn-demo}, top left, blue circles).
However, aggregate differences in entropy conflate the contributions of repertoire size, activity, and visit heterogeneity.
To disentangle these effects, we pair individuals with similar $N$ and $\hat{k}$ using nearest-neighbour matching~\cite{stuart-2010-matching}, then compute the mean relative entropy difference within matched male--female pairs ($\frac{X_{\text{Male}} - X_{\text{Female}}}{(X_{\text{Male}} + X_{\text{Female}})/2}$) and contrast the result against a permutation test in which gender labels are randomly reassigned (Methods:~\nameref{sec:methods-match}).

Controlling for repertoire size and activity reduces the observed gender gap from $5.16\%$ to $0.95\%$ (Fig.~\ref{fig:sn-demo}a, top left, green squares), indicating that the aggregate difference is driven primarily by variation in $N$ and $\hat{k}$ rather than by differences in visit heterogeneity.

Yet even when controlling for $N$ and $\hat{k}$, comparing males and females across the full population may be misleading, since the relationship between entropy and visit heterogeneity varies with $\hat{k}$ (see Limitation~1).
We therefore stratify individuals into three groups based on the combined deciles for $N$ and $\hat{k}$:
Inactive (1st--3rd deciles of both $N$ and $\hat{k}$), Moderately Active (4th–7th deciles), and Active (8th and above deciles).
These groups capture the majority of the sample, given the strong correlation between $N$ and $\hat{k}$ ($\rho=0.72$).
Within each group, we re-assess gender differences using matched pairs (Methods:~\nameref{sec:methods-match}).
When stratified in this way (Fig.~\ref{fig:sn-demo}, top right), entropy is higher for women only among the most active subgroup. This indicates that women distribute their visits more evenly across locations than men, but only when sequences are long and span many locations.

Notably, women exhibit higher temporal entropy ($\hat{S}^{temp}$) than men, even after controlling for $N$ and $\hat{k}$ (see Supplementary Material~\ref{si:impgen-stemp}), suggesting that visitation entropy alone misses differences in how visit sequences are ordered --- a dimension better captured by network-based approaches~\cite{desojo-2026-womens}.

These findings help clarify a discrepancy previously noted in~\cite{desojo-2026-womens}: CDR-based studies (Call Data Records) report higher entropy for men~\cite{gauvin-2020-gender, yuan-2016-analyzing, contreras-2023-linking, bertocchi-2025-big}, while survey studies and GPS-based analyses suggest the opposite~\cite{psylla-2017-role}, even before controlling for activity and repertoire size. 
Since location repertoires vary with spatial resolution~\cite{vanhoof-2018-comparing}, CDRs may underestimate the number of unique locations visited by women~\cite{yuan-2016-analyzing, gauvin-2020-gender, reisch-2021-behavioral, bertocchi-2025-big}, who tend to travel shorter distances~\cite{lenormand-2015-influence, yuan-2016-analyzing, gauvin-2020-gender, reisch-2021-behavioral}.
Consistent with this, we find that women make 12.98\% more visits than men at distances below 4~km (Supplementary Material~\ref{si:impgen-reso}).
This suggests that low-spatial-resolution datasets may selectively undercount women's short-distance visits, compressing their estimated repertoire size and biasing their measured entropy downward relative to men.

\textbf{Beyond gender: commuters and urbanization.}
The patterns identified for gender extend to other sociodemographic comparisons.
For commuters and non-commuters, the confounding effects of activity and repertoire size are even more pronounced: non-commuters exhibit larger repertoires, making them appear more entropic than commuters when unmatched (Fig.~\ref{fig:sn-demo}, center left, blue circles). Once activity and repertoire size are controlled for, this aggregate difference becomes negligible (Fig.~\ref{fig:sn-demo}, center left, green squares).
Stratifying by activity group, however, reverses the direction: among inactive individuals, commuters show higher entropy than non-commuters, whereas among the most active, the opposite holds (Fig.~\ref{fig:sn-demo}, center left, green squares; center right).

Comparing users from urban vs. rural areas further illustrates these limitations: without controls, urban residents appear to have a broader distribution of visits (on average, higher values of entropy); yet after controlling for $N$ and $\hat{k}$, these differences shrink substantially (Fig.~\ref{fig:sn-demo}, bottom left), and remain significant only in the moderately active group (Fig.~\ref{fig:sn-demo}, bottom right).

\subsection*{Discussion}
Our results reveal that visitation entropy, despite its widespread use as a measure of visit diversity, largely reflects how many locations individuals visit and how long their trajectories are --- two factors that together explain 90.7\% of its variance.
The dependence on repertoire size is a consequence of Shannon entropy's definition as a measure of compressibility, but we show that the normalization approaches often used in the literature, including dividing by $\log(k)$, fail to correct for it.
Sequence length introduces an additional dependence.
Entropy is a property of the underlying distribution and should theoretically be independent of sample size.
Yet, typical monthly sequences are too short to reliably capture the effect of low-probability locations in the visit distribution. 
We show that empirical estimates of $\hat{S}$ are systematically biased from the true entropy $S$, converging only when trajectories are long.
In experiments based on synthetic trajectories with realistic features, accurate estimation requires trajectories including more than roughly 600 visits.
Mobility behavior is therefore too rich to be captured by a single number: repertoire size, sequence length, and visit heterogeneity carry independent information that visitation entropy cannot disentangle.

These dependencies have direct implications for group comparisons.
Women exhibit higher entropy than men overall, but once activity and repertoire size are controlled for, this difference drops substantially.
The dependence on repertoire size helps clarify the discrepancy in gender differences in entropy between CDR and GPS-based studies~\cite{desojo-2026-womens}: CDRs may underestimate the number of unique locations visited by women~\cite{yuan-2016-analyzing, gauvin-2020-gender, reisch-2021-behavioral, bertocchi-2025-big}, who tend to travel shorter distances~\cite{lenormand-2015-influence, yuan-2016-analyzing}, consistent with women making 12.98\% more visits than men below 4km (Supplementary Material~\ref{si:impgen-reso}).

Similar patterns arise for commuters versus non-commuters and urban versus rural residents, where uncontrolled entropy differences shrink or reverse once repertoire size and activity are accounted for.
More broadly, we suggest these limitations extend beyond sociodemographic comparisons: changes in entropy over time may reflect the inclusion of new locations rather than genuine shifts in visit distributions, risking misattribution of behavioral change.

Finally, we show that controlling for $N$ and $\hat{k}$ through matching and stratification effectively isolates visit heterogeneity, and that individual mobility network metrics recover structural features of movement that entropy cannot capture.

In conclusion, we find that visitation entropy is a valuable descriptor of mobility, but it should be used with caution.
When applied without appropriate controls, it risks conflating different effects and may obscure comparisons across groups or over time.
Interpreting entropy alongside network metrics, controlling for unique visited locations, and ensuring trajectories are sufficiently long and well-sampled, provides a more reliable basis for comparing mobility patterns.


\section*{Methods}

\subsection*{Data description}\label{sec:meth-data}
This dataset uses mobility data collected between 2017 and 2019, previously analyzed in~\cite{alessandretti-2020-scales}, where stop locations were identified using the scalable stop-location detection algorithm InfoStop~\cite{aslak-2020-infostop}.
The stop location data captures the mobility patterns of 543,155 users.
To preserve privacy, spatial coordinates have been removed.
All users included had at least four months of data with positions recorded on at least 80\% of days. To ensure equal representation, we analyzed monthly sequences and sampled three months per user.

Individuals are located across ten countries (Japan, the United Kingdom, Germany, France, Spain, Taiwan, the Netherlands, Australia, the United States, and Sweden), are between 25 and 65 years old, and 44\% are female.
During app registration, participants self-reported their year of birth and gender (limited to “male” or “female”).
We acknowledge that this binary classification limits the scope of gender analysis in this study.
Accordingly, we use the terms “woman” and “female,” as well as “man” and “male,” interchangeably throughout the analysis, while acknowledging that no universally accepted correspondence exists between these terms and an individual’s sex or gender identity~\cite{hawkesworth-2013-sex, ritz-2024-we}.
The home country is defined as the country where each user spent the most time, based on their stop sequences. Home coordinates are then linked to the 2019 Global Human Settlement Layer (GHS)~\cite{florczyk-2019-ghsl}, which classifies areas as urban, suburban, or rural using a 1km resolution grid. Finally, commuting status is estimated by identifying individuals’ home and work locations with the state-of-the-art HoWDe algorithm (see section~\nameref{sec:methods-homework} for additional details)~\cite{desojo-2025-establishing}.

\subsection*{Preferential Return}\label{sec:methods-epr}
We consider synthetic mobility sequences generated using the Exploration and Preferential Return (EPR) model, a well-established framework in human mobility~\cite{song-2010-modelling}.
Following the literature, at equilibrium, a mobility trajectory can be generated using the preferential return mechanism:
(i) An individual knows $k$ distinct locations. These $k$ locations are ranked from the most to the least likely to be visited;
(ii) Over a given time span, the individual makes $N$ visits to these locations;
(iii) The probability to pick a location with rank $i$ follows a power-law distribution $p(i) = \frac{i^{-\alpha}}{\zeta_{\hat{k}}(\alpha)}$, where $\alpha$ is the power-law exponent, and $\zeta_{k}(\alpha) = \sum_{i=1}^{k} i^{-\alpha}$ is the generalized harmonic number (a truncated form of the Riemann zeta function), which ensures normalization.

Each individual is thus characterized by three parameters: the maximum size of their \emph{repertoire} of locations ($k$), the \emph{heterogeneity} in visit distribution across locations ($\alpha$), and their \emph{activity} level ($N$)---or sequence length.

The model generates a sequence of $N$ visits to $\hat{k}\leq k$ distinct locations, where each of these locations $i$ appears in the sequence with frequency $f(i)$.
The empirical entropy is then computed from the visit frequencies $f(i)$ as
$$\hat{S} = -\sum_{i=1}^{\hat{k}} f(i) \log_2 f(i),$$
while the entropy of the generative model can be derived analytically as
$$S = -\sum_{i=1}^{k} p(i) \log_2 p(i) = \frac{\alpha}{\zeta_k(\alpha)} \sum_{i=1}^{k} i^{-\alpha} \log_2 i + \log_2 \zeta_{k}(\alpha),$$
where $p(i)$ is the power-law distribution introduced above.

Similarly, we can estimate the empirical $\hat{\alpha}$ of the sequences by applying maximum likelihood estimation (MLE) over the simulated visit distributions.
This involves optimising the power-law log-likelihood to recover the best-fit exponent.
This setup provides a controlled benchmark, allowing us to disentangle the effects of repertoire size, activity, and heterogeneity on entropy estimates.

\subsection*{Entropy regressions}\label{sec:meth-regression}

We used gradient-boosted decision trees (XGBoost) to evaluate the extent to which entropy can be attributed to simple activity measures and subsequently to the structural properties of mobility networks.
Models were trained to predict individual visitation entropy using progressively richer sets of predictors:
(i) activity level ($N$) and repertoire size ($k$);
(ii) the full set of network metrics, including: degree centrality of the top three nodes (1st, 2nd, and 3rd highest degree), average clustering coefficient, number of cycles, cycle length, network diameter, transitivity, and node connectivity.

All models were implemented without hyperparameter tuning. We trained on 80\% of the data and evaluated performance on the remaining 20\%. Model performance was quantified using the coefficient of determination ($R^2$), averaged across 5-fold cross-validation. Feature importance was assessed using SHAP values, which measure the marginal contribution of each predictor to the model’s output.

To test whether structural features become more relevant among highly active users, we repeated the analysis within strata defined by combined deciles of activity and repertoire size, classifying individuals into inactive, moderate, and active groups (see Supplementary Materials~\ref{si:lim3-xgboost}).

\subsection*{Individual mobility networks}\label{sec:methods-netw}
To examine what entropy can capture beyond N and k, we represent each individual’s one-month sequence of visited locations as a directed, weighted network.
Nodes correspond to unique locations, and a directed edge $i \rightarrow j$ indicates that the individual visited location $j$ immediately after location $i$. The weight $w_{ij}$ of each edge counts the number of such transitions over the observation window (one month).

The number of nodes, $k$, reflects the repertoire of unique locations visited.
Total activity, $N$, is defined as the sequence length, i.e. the number of visits within the observation window.
Formally, $N = \sum_{(i,j) \in E} w_{ij} + 1,$ where $E$ is the set of directed edges and $w_{ij}$ their weights.
The $+1$ accounts for the fact that a sequence of $N$ visits yields $N-1$ transitions. Because all sequences span the same time window, $N$ is directly comparable across individuals.

The number of visits to a specific node can be approximated by summing the weights of its incident edges and dividing by two, since each visit typically involves both an arrival and a departure.
This approximation holds for most nodes but fails at the sequence boundaries: the first and last nodes may involve only a single connection, leading to an underestimation of their true visit count.
For example, in the sequence $[1,2,1,2,1]$, the total activity is $N=5$, which equals the sum of edge weights ($w_{12}+w_{21}=4$) plus one. Node $1$ is visited three times, not twice as predicted by the degree-based approximation, while node $2$ is visited exactly twice.

In practice, these discrepancies are minor, and node visit counts remain approximately proportional to their degree divided by two. This representation offers a simple and interpretable mapping between the network structure and individual mobility behaviour.

\subsection*{Network metrics}\label{sec:meth-nwmet}
We quantify structural properties of individual mobility networks using standard network metrics.
First, we measure the \emph{number of cycles}, defined as the count of simple closed paths in the network (e.g., A--B--C--A).
Cycles capture recurrent multi-stop patterns beyond back-and-forth travel.

Second, we compute the \emph{average clustering coefficient}, which measures the tendency of visited locations to form tightly connected subgroups. For a node $i$ with degree $k_i$, the local clustering coefficient is
\[
    C_i = \frac{2e_i}{k_i(k_i - 1)},
\]
where $e_i$ is the number of edges between the neighbours of $i$. The network-level clustering coefficient is then obtained as
\[
    C = \frac{1}{|V|} \sum_{i \in V} C_i,
\]
where $|V|$ denotes the number of nodes.  

Finally, we assess \emph{degree centrality}, which quantifies the relative importance of each node. The degree centrality of a node $i$ is defined as
\[
    d_i = \frac{k_i}{|V|-1},
\]
where $k_i$ is the degree of node $i$. We compute degree centrality across all nodes and report values for the top one, two, and three most central nodes.
These typically correspond to home, work, and a third key location~\cite{desojo-2026-womens}.
For the regression analyses, we additionally include: \textit{Network density}, ratio of observed edges to the maximum possible (in a fully connected network), reflecting overall connectivity; \textit{Transitivity}, fraction of closed triplets over all triplets, measuring triadic closure; \textit{Average cycle length}, mean number of nodes per cycle, capturing the typical length of recurrent travel loops; \textit{Network diameter}, the longest shortest path between any two nodes, describing the span of the mobility network; \textit{Average node connectivity}, expected number of nodes that must be removed to disconnect a random pair, reflecting network robustness.

\subsection*{Partial correlation}\label{sec:methods-pcor}
To account for the influence of repertoire size ($k$), we report partial correlations between visitation entropy ($X$) and each network metric ($Y$), controlling for $k$: $\rho_{X,Y\cdot k}$.
Formally, this is equivalent to computing the Pearson correlation between the residuals $e_X$ and $e_Y$ obtained from linear regressions of $X$ on $k$ and $Y$ on $k$, respectively.
This approach isolates the association between entropy and structural network properties that is not explained by repertoire size.

\subsection*{Temporal Entropy}\label{sec:methods-realS}
To capture the temporal dimension of mobility, we compute the \emph{temporal entropy}, $S^{temp}$, using the Lempel--Ziv estimator~\cite{song-2010-limits}.
Unlike visitation entropy, which depends only on the frequency distribution of visited locations, the temporal entropy incorporates the full spatiotemporal order of a trajectory.
Formally, following Song et al.~\cite{song-2010-limits}, let $T_i = \{X_1, X_2, \dots, X_L\}$ denote the time-ordered sequence of locations visited by user $i$ across consecutive intervals.
The temporal entropy $S^{temp}_i$ is defined as
\[
    S^{temp}_i = - \sum_{T'_i \subset T_i} P(T'_i) \log_2 P(T'_i),
\]
where $P(T'_i)$ is the probability of observing a particular subsequence $T'_i$ in the trajectory $T_i$.
By construction, $S_i$ depends not only on the frequency of visits, but also on the order in which locations are visited and the time spent at each, thus reflecting the full spatiotemporal structure of mobility.

For each user, the following inequality holds: $S^{temp}_i \leq S_i \leq S^{rand}_i$,
where $S_i$ is the uncorrelated (visitation) entropy and $S^{rand}_i = \log_2(k)$ is the random entropy, i.e. the maximum possible entropy given the repertoire size $k$.

The Lempel--Ziv estimator provides a practical approximation of $S^{temp}_i$ from finite sequences by exploiting the compressibility of the trajectory: regular sequences yield shorter encodings, while irregular sequences require longer code lengths~\cite{jensen-2010-estimating}.

\subsection*{Nearest neighbor matching}\label{sec:methods-match}
To overcome the limitations of visitation entropy and isolate differences in visit heterogeneity from activity and repertoire size, we apply a nearest-neighbour matching procedure~\cite{stuart-2010-matching}.
For each comparison, individuals in one demographic group are matched to their closest counterpart in the other group based on (i) total number of visits ($N$) and (ii) repertoire size ($k$).
Both covariates are standardised before matching. We use a 1-nearest neighbour algorithm with Euclidean distance, ensuring that each individual is paired with the most similar counterpart in covariate space.

Matching is carried out both across the full population and within subgroups defined by activity level (inactive, moderate, active). To ensure comparability, we restrict the allowable difference in repertoire size between pairs. For the overall sample, we require nearly exact matching ($\Delta k \leq 1$).
Within subgroups, we relax this threshold slightly to reflect natural variability: up to one location for inactive users, two for moderate users, and four for active users---corresponding roughly to the 10th percentile of repertoire size in each group.

After matching, we compute the average relative difference in visitation entropy between activity groups.
To evaluate significance, we compare these values to a null distribution obtained by randomly permuting group labels while keeping the full matching pipeline intact.
This randomisation test provides a baseline against which observed sociodemographic differences can be assessed.

\subsection*{Home and work detection for commuting status classification}\label{sec:methods-homework}
We detect home and work locations using \textit{HoWDe}, a validated algorithm for longitudinal GPS data~\cite{desojo-2025-howde}.
HoWDe integrates multiple heuristics---including visit duration, temporal regularity, and frequency of occurrence—to identify Home and Work locations.
The algorithm is designed to be robust to irregular travel behaviour and data sparsity. Validation against ground-truth datasets demonstrates high accuracy across demographic groups, settlement types, and countries (see~\cite{desojo-2025-howde}).

Commuting status is derived from the presence of a work location. A user is classified as a commuter if a work location is detected consistently for at least one month.
If users have secondary locations but none meet the frequency threshold defined by HoWDe for a work location, they are classified as non-commuters.
Users with no detectable work location, due to insufficient or missing data, are left unlabelled.

\subsection*{Urbanization classification}\label{sec:methods-urb}
The level of urbanization surrounding each individual’s home location was determined using the 2019 GHS Settlement Model grid at 1 km resolution~\cite{florczyk-2019-ghsl}.
The GHS framework classifies settlement typologies into urban, suburban, and rural based on population size, density, and built-up area. For this analysis, we only consider the urban and rural categories.

\section*{Code and data availability}
Data required to ensure replicability of the analyses conducted in this study will be available at: \href{https://doi.org/10.11583/DTU.32771463}{DOI: 10.11583/DTU.32771463}.
Code to reproduce all analyses is openly available at: \href{https://github.com/sdesojo/limits-visit-entropy}{github.com/sdesojo/limits-visit-entropy}.
All data processing and analysis were carried out in compliance with the European Union’s General Data Protection Regulation (GDPR; Regulation 2016/679).

\section*{Acknowledgments}
L.A. and S.D.S. acknowledge support from the Independent Research Fund Denmark (DFF) through the DFF-Research Project 1 (Inge Lehmann) grant ``Gender Gaps in Human Mobility''.

\section*{Author contributions}
Conceptualization: SDS, SL, LA.
Methodology: SDS, LA.
Investigation: SDS.
Visualization: SDS.
Funding acquisition: LA.
Project administration: LA.
Supervision: LA, SL.
Writing -- original draft: SDS, LA.
Writing -- review \& editing: SDS, SL, LA.

\section*{Competing interests} Authors declare that they have no competing interests.

\section*{Materials and correspondence} Correspondence and material requests should be addressed to Sílvia de Sojo, e-mail: \href{mailto:sdsc@dtu.dk}{sdsc@dtu.dk}

\newpage
\bibliographystyle{unsrtnat}
\bibliography{references-entropy}

\newpage
\setcounter{section}{0}
\setcounter{figure}{0}
\renewcommand{\thesection}{S\arabic{section}}
\renewcommand{\thefigure}{S\arabic{figure}}

\begin{center}
    {\Large\textbf{Supplementary Material}}\\[6pt]
    {\large\textit{The limits of visitation entropy as a summary of mobility patterns}}
\end{center}

\vspace{1em}

\section{Numerical validations of the simulation framework}\label{si:numval}
To verify the consistency between simulations and analytical expectations, we compared empirical estimates from the Preferential Return (EPR) model with their corresponding analytical expressions.

\begin{figure}[H]
    \centering
    \includegraphics[width=1\linewidth]{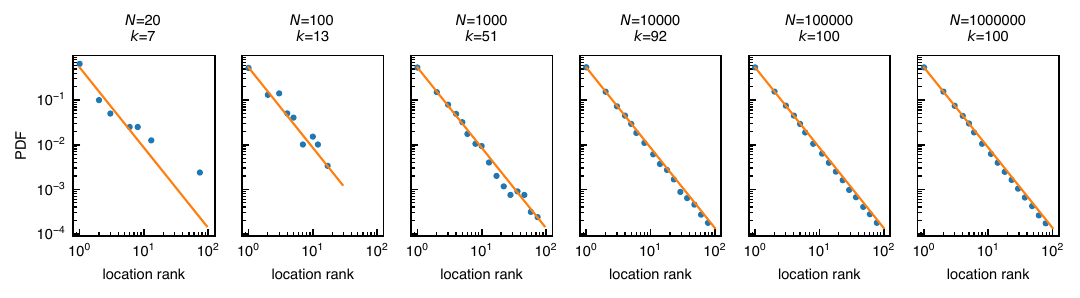}
    \caption{Validation of simulated visit frequencies against the theoretical distribution. Simulated visit probabilities (blue dots) compared to the analytical distribution (orange line) for $\hat{\alpha} = 1.8$ and $\hat{k} = 100$, across increasing sequence lengths $N$. As $N$ grows, the simulated distributions converge to the analytical expectation, and the observed repertoire size $\hat{k}$ approaches the theoretical maximum.}
    \label{fig:si-numval1}
\end{figure}

First, we tested whether simulated visit frequencies match the theoretical distribution.
Fixing $\hat{\alpha} = 1.8$ and $\hat{k} = 100$, we generated sequences with varying lengths $N$. As expected, the agreement between simulations and the analytical distribution improves with increasing $N$ (Fig.~\ref{fig:si-numval1}).

\begin{figure}[H]
    \centering
    \includegraphics[width=1\linewidth]{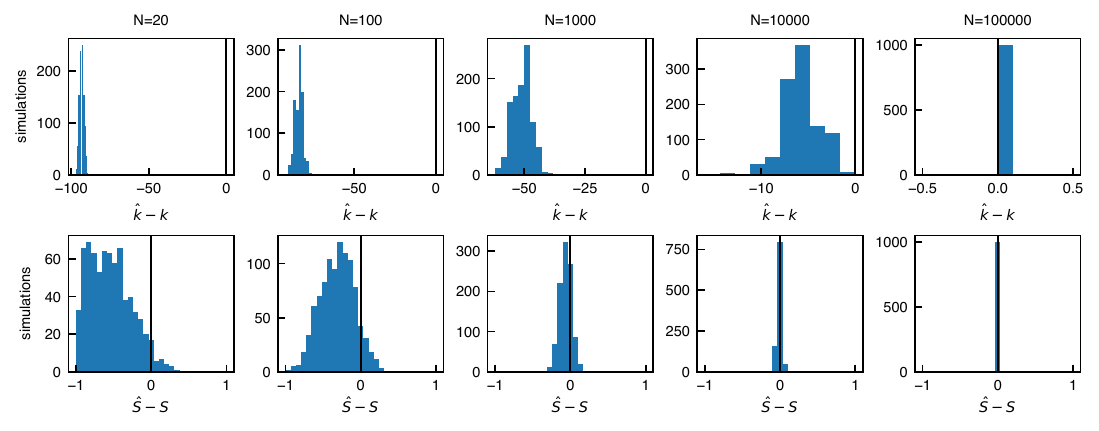}
    \caption{Finite-sample effects on repertoire size and entropy estimates. Distribution of differences between simulated and analytical values of repertoire size ($\hat{k} - k$, top) and entropy ($\hat{S} - S$, bottom) for $\hat{\alpha} = 1.8$, $\hat{k} = 100$, and varying sequence lengths $N$. For small $N$, the repertoire size is underestimated, leading to systematic downward bias in entropy. As $N$ increases, both $\hat{k}$ and $\hat{S}$ converge to their analytical values.}
    \label{fig:si-numval2}
\end{figure}

We next examined the entropy of simulated sequences.
Using the same parameters ($\hat{\alpha} = 1.8$, $\hat{k} = 100$), we compared empirical entropy $\hat{S}$ with the analytical entropy $S$.
For small $N$, the simulated repertoire size $\hat{k}$ falls short of the theoretical maximum $k$, leading to a systematic underestimation of entropy.
The estimates converge to the analytical values only when $N$ is sufficiently large (Fig.~\ref{fig:si-numval2}).

In summary, for small $N$, $k$ is underestimated, resulting in a systematic downward bias in entropy.
Only for large $N$ do the inferred parameters and entropy converge to their true analytical values. As a consequence, for small $N$, we underestimate $S$.

\section{Limitation 1: Repertoire size modulates entropy's sensitivity to the visit distribution heterogeneity}\label{si:Ssens}
When analytically examining the joint role of repertoire size $k$ and heterogeneity $\alpha$ in shaping entropy $S$, we found that the sensitivity of entropy to changes in $\alpha$ depends on repertoire size.

\begin{figure}[H]
    \centering
    \includegraphics[width=1\linewidth]{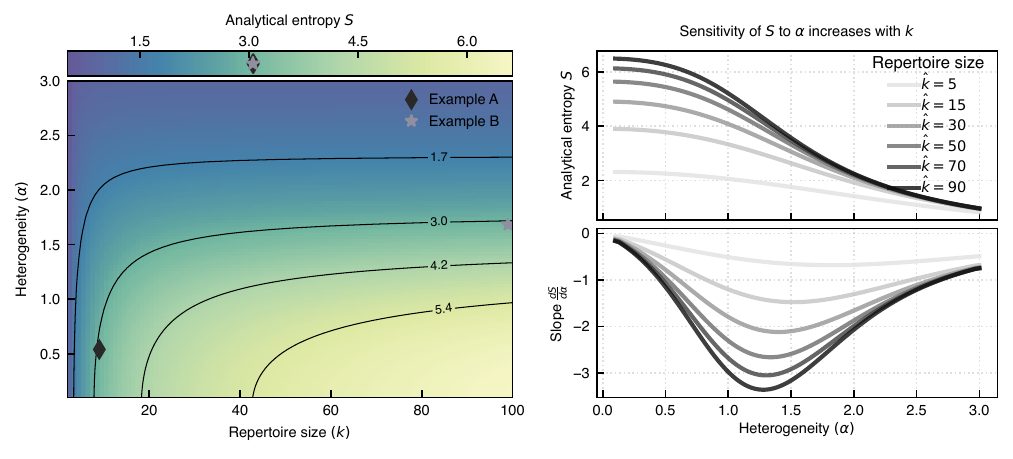}
    \caption{\textbf{Repertoire size modulates entropy's sensitivity to the visit distribution heterogeneity.} Left: Analytical entropy ($S$, colour scale) as a function of heterogeneity ($\alpha$) and repertoire size ($k$), showing that the same entropy value can emerge from distinct behavioural patterns. Right: Analytical visitation entropy $S$ as a function of heterogeneity $\alpha$ for different repertoire sizes $k$ (top), and the corresponding slopes $\tfrac{dS}{d\alpha}$ (bottom). Larger repertoires amplify the decline of $S$ with $\alpha$, indicating that entropy becomes increasingly sensitive to changes in heterogeneity as $k$ grows.}
    \label{fig:si-lim1-salpha}
\end{figure}

This effect is particularly clear when exploring parameter ranges comparable to empirical data, where the median repertoire size is $k = 19$.
Here we illustrate this relationship further. Figure~\ref{fig:si-lim1-salpha} shows the analytical visitation entropy $S$ across values of $\alpha \in [0.1, 3]$ and repertoire sizes $k \in [2, 90]$.
The dependence is reflected in the increasingly steep negative slope of $S$ with respect to $\alpha$ at larger $k$ (Fig.~\ref{fig:si-lim1-salpha}, left).

\begin{figure}[H]
    \centering
    \includegraphics[width=1\linewidth]{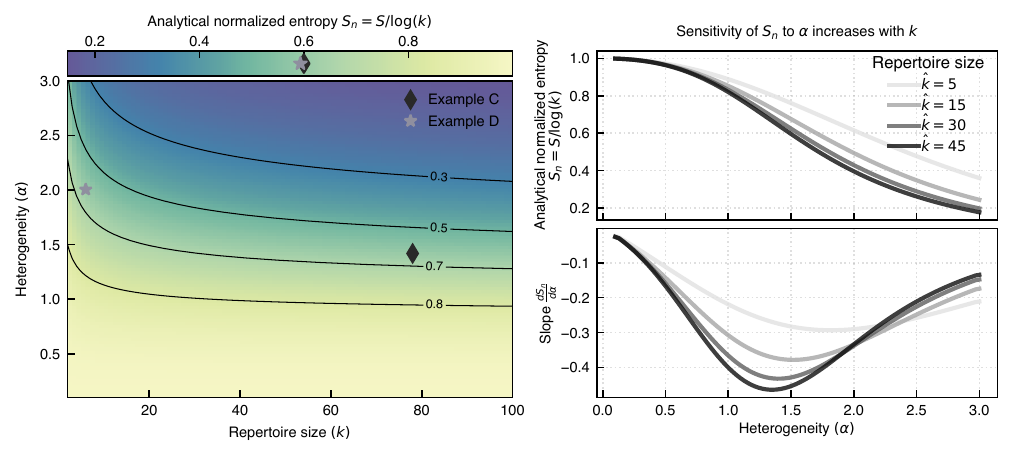}
    \caption{\textbf{Repertoire size modulates the sensitivity of normalized entropy to the visit distribution heterogeneity.} Left: Normalized entropy ($S_n = \tfrac{S}{\log_2(k)}$, colour scale) as a function of heterogeneity ($\alpha$) and repertoire size ($k$), showing that identical values of $S_n$ can arise from distinct behavioral patterns. Right: Normalized entropy $S_n$ as a function of heterogeneity $\alpha$ for different repertoire sizes $k$ (top), and the corresponding slopes $\tfrac{dS_n}{d\alpha}$ (bottom). Although normalization reduces absolute differences across $k$, larger repertoires still amplify the decline of $S_n$ with $\alpha$, indicating that sensitivity to heterogeneity is not removed by normalization.}
    \label{fig:si_lim1_Snormsens}
\end{figure}

Importantly, this dependence persists even after normalizing entropy,
$S_n = \frac{S}{\log_2(k)}$,
indicating that normalization does not fully remove the influence of repertoire size (Fig.~\ref{fig:si_lim1_Snormsens}).

\section{Limitation 2: Additional configurations for the finite-sample analysis}\label{si:lim2-otherconfigs}
In Limitation 2, we examined the impact of sequence length $N$ on entropy estimates under fixed parameters $\alpha = 1.65$ and $k = 60$, yielding $S = 3.01$.
Here, we extend the analysis to additional parameter configurations to test the robustness of these findings.

\begin{figure}[H]
    \centering
    \includegraphics[width=1\linewidth]{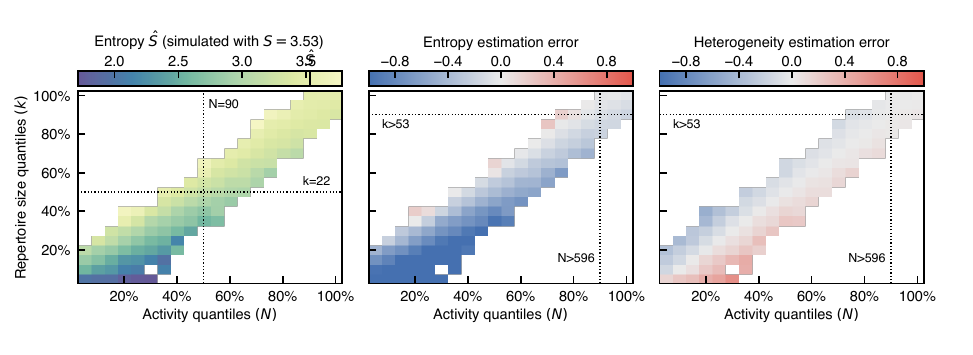}
    \caption{\textbf{Finite-sample effects for larger repertoire sizes and more even distributions.}
        Left: Entropy from synthetic traces generated with heterogeneity $\alpha = 1.5$, repertoire size $k = 80$, resulting in analytical entropy $S = 3.53$, for increasing $N$.
        Centre: Entropy estimation error ($\hat{S} - S$) in synthetic traces.
        Right: Heterogeneity estimation error ($\hat{\alpha} - \alpha$), showing that only for large $N$ and $k$, the estimation error in heterogeneity converges toward the true value $\alpha$.}
    \label{fig:si-lim2a}
\end{figure}

First, we consider a broader repertoire and a flatter (less heterogeneous) visit distribution, with $k = 80$ and $\alpha = 1.5$, for which the analytical entropy increases to $S = 3.53$. As in the baseline case, empirical entropy systematically underestimates the analytical value for small $N$, converging only when trajectories are sufficiently long (Fig.~\ref{fig:si-lim2a}, left and centre).

As $N$ increases, the heterogeneity estimation error ($\hat{\alpha} - \alpha$) converges toward the true value $\alpha$ (Fig.~\ref{fig:si-lim2a}, right; see Methods for details on estimating $\alpha$).
For small repertoire sizes, short sequences tend to underestimate $\alpha$, yielding distributions that appear flatter than they actually are. With larger $N$, the bias diminishes, and the estimated distribution more closely reflects the underlying one.
However, beyond a certain threshold, further increases in $N$ lead to an overestimation of $\alpha$.
This behavior illustrates how statistical fluctuations in the distribution of visits, driven by the difficulty of capturing low-probability locations, systematically distort entropy estimates.

\begin{figure}[H]
    \centering
    \includegraphics[width=1\linewidth]{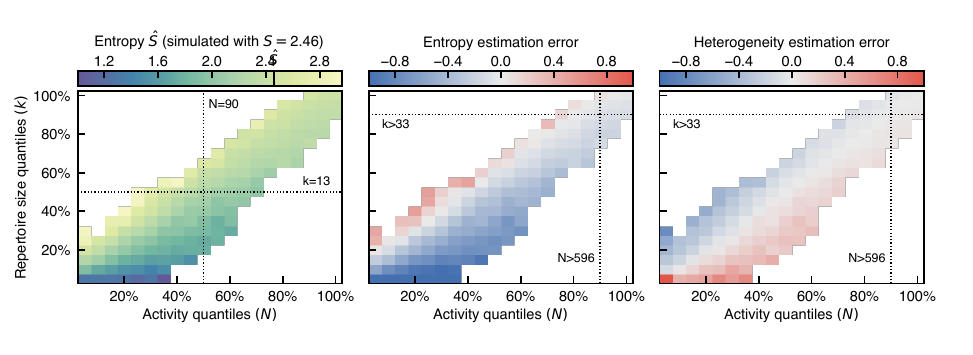}
    \caption{\textbf{Finite-sample effects for larger repertoire sizes and less even distributions.} Left: Entropy from synthetic traces generated with heterogeneity $\alpha = 1.9$, repertoire size $k = 80$, and analytical entropy $S = 2.46$, across increasing $N$. Centre: Entropy estimation error ($\hat{S} - S$) in synthetic traces. Right: Heterogeneity estimation error ($\hat{\alpha} - \alpha$), showing that short sequences overestimate $\alpha$ while convergence toward the true value requires large $N$.}
    \label{fig:si-lim2b}
\end{figure}

With the same repertoire size but for more peaked distributions ($k = 80$, $\alpha = 1.9$), the analytical entropy decreases to $S = 2.46$. As in the previous case, statistical fluctuations persist: short sequences, with a limited number of states (small repertoire size), fail to capture the underlying heterogeneity of the model and produce systematically biased entropy estimates (Fig.~\ref{fig:si-lim2b}).

\begin{figure}[H]
    \centering
    \includegraphics[width=1\linewidth]{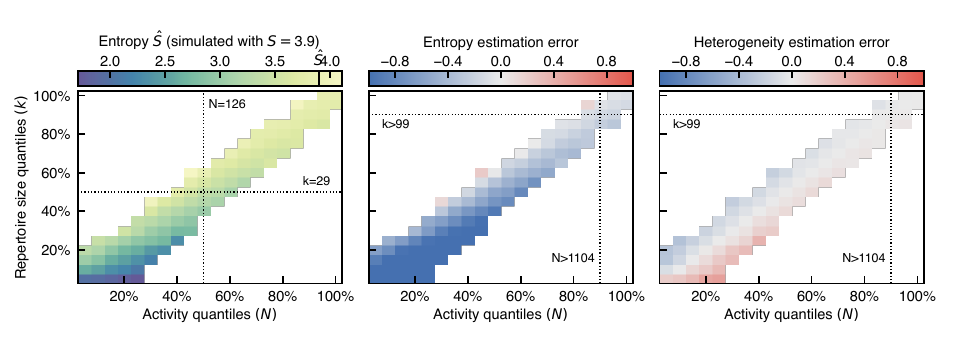}
    \caption{\textbf{Finite-sample effects for yearly datasets.}
        Left: Entropy from synthetic traces generated with heterogeneity $\alpha = 1.5$, repertoire size $k = 240$, and analytical entropy $S = 3.9$, across increasing $N$.
        Centre: Entropy estimation error ($\hat{S} - S$) in synthetic traces.
        Right: Heterogeneity estimation error ($\hat{\alpha} - \alpha$), reinforcing the finite-sample bias.}
    \label{fig:si-lim2-k240a15}
\end{figure}

Finally, to reproduce the scale of a yearly dataset, we tested simulations up to $N = 2000$, a larger repertoire size of $k = 240$ with $\alpha = 1.5$, resulting in $S = 3.9$.
Now, entropy estimates are consistently biased downward for small $N$, and again approach the analytical value for long sequences (Fig.~\ref{fig:si-lim2-k240a15}).

Across all tested configurations, the finite-sample bias drives statistical fluctuations, systematically biasing the empirical entropy when $N$ is small, independent of the analytical repertoire size or the visit distribution heterogeneity.
This result underscores that reliable entropy estimates require very long sequences with many observed states --- a condition rarely met in typical mobility studies.

\section{Limitation 2: Validation across normalization approaches}\label{si:lim2-othernorm}

We demonstrated that finite-sample effects result in a systematic bias of entropy when the number of observations, $N$, is of the same order as the repertoire size, $k$.
Here, we test whether this bias persists under common normalization and correction schemes.

\begin{figure}[H]
    \centering
    \includegraphics[width=1\linewidth]{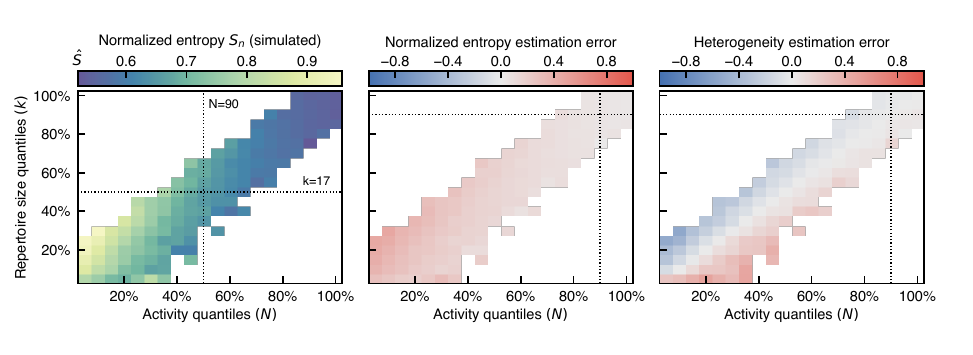}
    \caption{\textbf{Finite-sample effects under normalization.}
        Left: Normalised entropy $S / \log_2(k)$ from synthetic traces with $\hat{\alpha} = 1.65$ and $\hat{k} = 60$ (analytical $\hat{S} = 3.01$), across increasing $N$.
        Centre: Estimation error in entropy ($\hat{S} - S$).
        Right: Estimation error in heterogeneity ($\hat{\alpha} - \alpha$), showing the same pattern of overestimation at small $N$ and underestimation at large $N$.
        Normalisation by $\log_2(k)$ systematically biases entropy upward relative to the analytical value.}
    \label{fig:si-lim2-norm}
\end{figure}

First, we consider the normalized entropy $S / \log_2(k)$.
Figure~\ref{fig:si-lim2-norm} shows that, although normalization rescales the estimates, it systematically biases entropy upwards, especially for small $N$.
The estimated heterogeneity $\alpha$ is underestimated for short sequences and overestimated for large $N$, consistent with finite-sample effects.

\begin{figure}[H]
    \centering
    \includegraphics[width=1\linewidth]{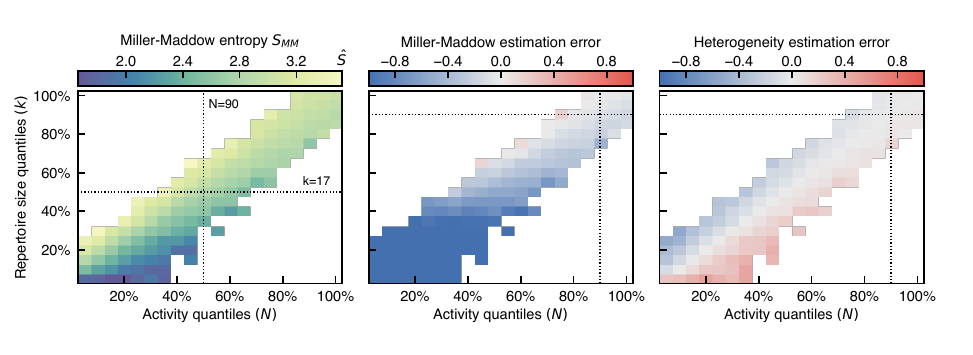}
    \caption{\textbf{Finite-sample effects under the Miller--Madow correction.}
        Left: Corrected entropy from synthetic traces with $\hat{\alpha} = 1.65$ and $\hat{k} = 60$ (analytical $\hat{S} = 3.01$), across increasing $N$.
        Centre: Estimation error in entropy ($\hat{S} - S$).
        Right: Estimation error in heterogeneity ($\hat{\alpha} - \alpha$), showing the same pattern as before.
        Despite the correction, entropy remains biased downward for small $N$.}
    \label{fig:si-lim2-mm}
\end{figure}

Next, we applied the Miller--Madow correction, which adjusts entropy estimates by adding $\frac{k-1}{2N}$ to account for sample size.
Figure~\ref{fig:si-lim2-mm} shows that entropy remains underestimated, especially for small $N$.
The same systematic bias in $\alpha$ persists, with overestimation at small $N$ and underestimation at large $N$ when repertoire sizes are small.

In summary, both normalization and Miller--Madow correction fail to eliminate the entropy biases.
Normalisation by $\log_2(k)$ systematically inflates entropy for short sequences, while Miller--Madow maintains the downward bias.

\section{Limitation 3: Entropy regressions}\label{si:lim3-xgboost}
\begin{figure}[H]
    \centering
    \includegraphics[width=\linewidth]{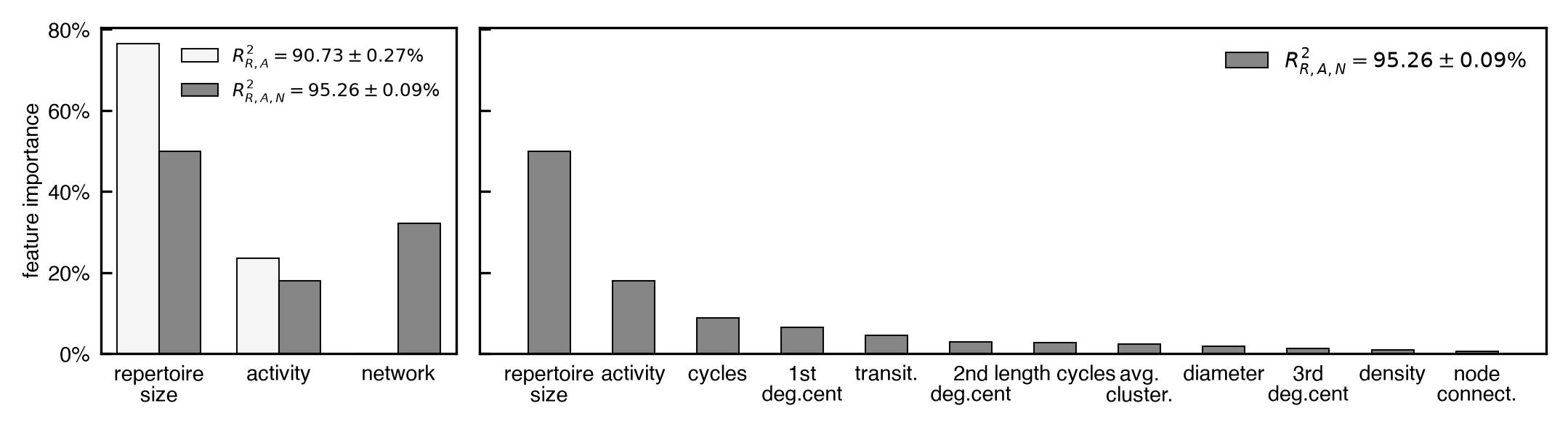}
    \caption{\textbf{Feature importance in entropy regressions.}
        Bar plots show the contribution of predictors to model performance based on SHAP values. The left panel compares models with and without network information: the baseline model using activity and repertoire size (RA, light gray) and the extended model including network features (RAN, dark gray). The right panel shows the detailed decomposition of feature importance within the network-only model. Reported values in the legend indicate mean $R^2$ and standard deviation across cross-validation folds.
        Here, the number of cycles is highly correlated with repertoire size (i.e., the total number of nodes in the network) and therefore does not fully capture the underlying network structure.}
    \label{fig:si-sn-xgboost-models-all}
\end{figure}

To validate the theoretical result that entropy is fundamentally driven by repertoire size ($k$) and activity level ($N$), we trained a random forest regression model (XGBoost) using only these two predictors.
Consistent with this expectation, the model explained $90.73\%\pm0.27$ of the variance in entropy across individuals in the empirical data, with repertoire size accounting for $76.5\%$ of total feature importance (see Fig.~\ref{fig:si-sn-xgboost-models-all}, left, light grey).
This result demonstrates that simple differences in activity and repertoire largely determine visitation entropy, rather than the heterogeneity or other structural aspects of mobility.

Next, we extended the models by incorporating network features as additional predictors alongside activity and repertoire size. This raised the explained variance to $95.26\%\pm0.09$ (see Figure~\ref{fig:si-sn-xgboost-models-all}, left), a modest gain of approximately 4.5 percentage points. Repertoire size remained the feature with the highest importance ($49.9\%$), followed by activity ($18\%$), number of cycles ($8.8\%$), and degree centrality of the top 1 ($6.5\%$) and top 2 ($2.9\%$) most central nodes.
However, interpreting the importance of individual network features is potentially misleading, given the strong correlations some features share with activity and repertoire size --- for example, the number of cycles.
Overall, the limited additional variance explained by network features is consistent with the view that structural aspects of mobility contribute marginally once $k$ and $N$ are accounted for.

\begin{figure}[H]
    \centering
    \includegraphics[width=0.8\linewidth]{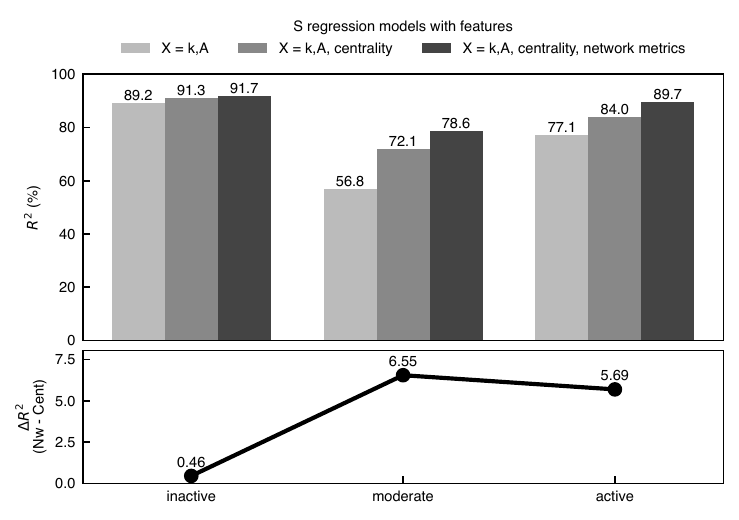}
    \caption{\textbf{Entropy regressions across activity groups.}
    Top: Explained variance ($R^2$) from models including different sets of predictors: activity and repertoire size (light grey); activity, repertoire size and degree centrality (grey); and all the previous plus additional network metrics (dark grey). Results are shown for three activity-level groups (from left to right): inactive, moderate, and active, defined by combined deciles of activity and repertoire size.
    Bottom: Marginal increase in $R^2$ when adding network metrics on top of centrality measures ($\Delta R^2 = R^2_{\text{Nw}} - R^2_{\text{Cent}}$), showing that structural network features provide only minor additional explanatory power, especially for less active users.}
    \label{fig:si-sn-xgboost-models-marginal}
\end{figure}

Finally, we validated these findings by testing performance across groups of increasing activity levels, defined by combined deciles of activity and repertoire size. This stratified analysis enables us to assess whether structural features contribute more strongly to highly active users. Figure~\ref{fig:si-sn-xgboost-models-marginal} shows that, compared to the baseline model (using activity and repertoire size only; light gray), incorporating all network features (black) increases the explained variance more substantially for individuals in the moderate and active groups.

To disentangle the contribution of centrality measures from other network properties, we trained an additional model including only activity, repertoire size, and the centrality of the top three nodes (gray).
We then assessed the marginal increase in explained variance when comparing this model to the one that also included broader network metrics (see Fig.~\ref{fig:si-sn-xgboost-models-marginal}).
The results show that the added contribution of network-level features is the lowest in the inactive group ($0.46$ percentage points increase), compared to the moderate and active groups (where network features contribute respectively to $6.55$ and $5.69$ percentage points).
This is consistent with the fact that network structure becomes less informative as the number of nodes decreases: for individuals with small repertoires, mobility networks are simple enough that $k$ and $N$ already capture most of the relevant variation in the entropy.

Together, these results confirm that entropy can be predicted almost entirely from activity and repertoire size. Structural properties of mobility networks provide only marginal additional explanatory power, and their contribution becomes relevant only among the most active users.

\section{Gender differences in travel distances}\label{si:impgen-reso}

\begin{figure}[H]
    \centering
    \includegraphics[width=1\linewidth]{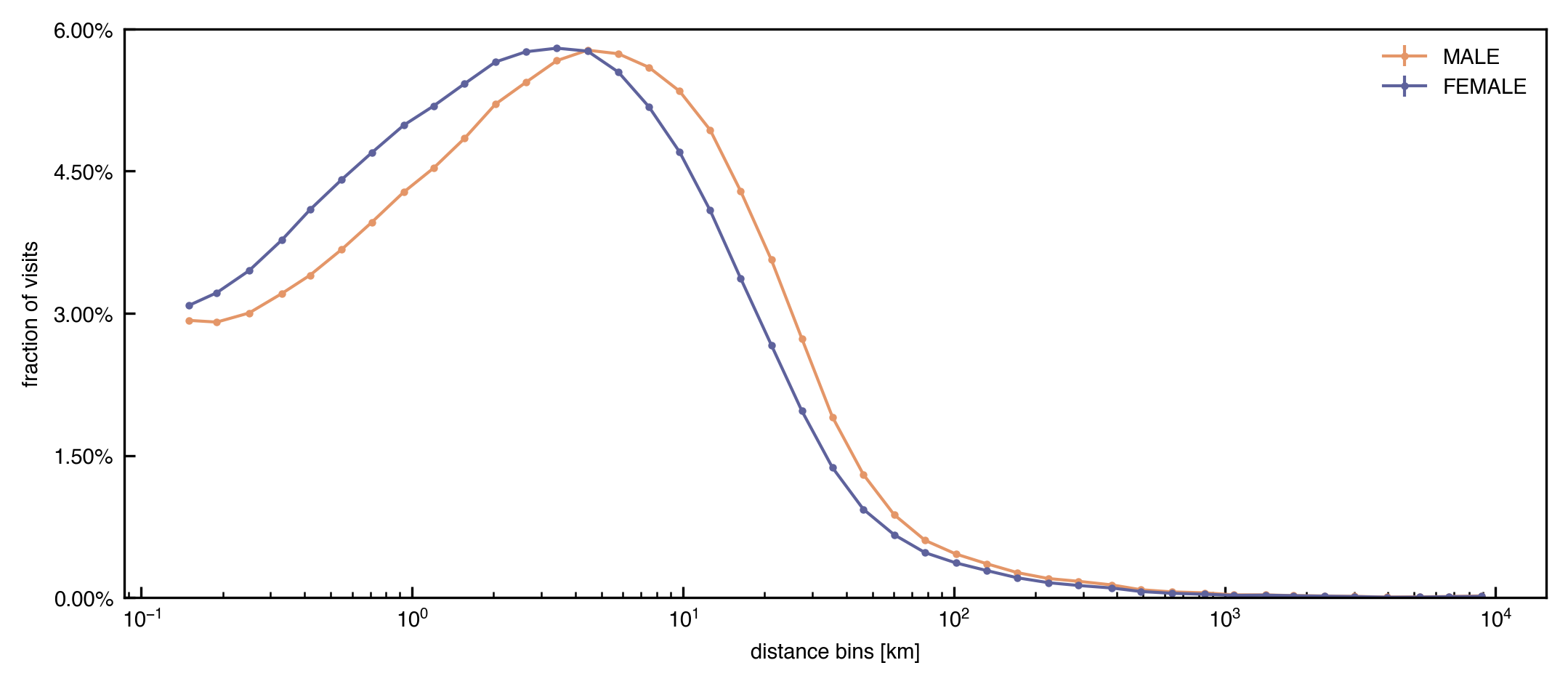}
    \caption{\textbf{Gender differences in travel distances.}
        Fraction of visits by trip distance bin for men (orange) and women (blue). Women make more short-range trips ($\leq 4$ km), while the distributions converge at medium and long distances.}
    \label{fig:si-reso}
\end{figure}

To examine gender differences across spatial scales, we computed the distribution of visits by distance between locations, binned separately for men and women.
Our results show that at short distances ($\leq 4$ km), women make $12.98\%$ more visits than men.
This excess declines with distance, and beyond $\sim 10$ km the distributions overlap almost entirely (Fig.~\ref{fig:si-reso}).

\section{Gender differences in temporal entropy}\label{si:impgen-stemp}

We use temporal entropy, $\hat{S}^{temp}$ (see Methods and Song et al., 2010), to capture aspects of mobility not reflected by visitation entropy.
Unlike visitation entropy, which considers only the frequency of visits to locations, temporal entropy incorporates the ordering of trajectories by accounting for the probability of observing time-ordered subsequences $T_i'$ within an individual's sequence of visits. This allows us to quantify regularities in how locations are sequenced over time.

\begin{figure}[H]
    \centering
    \includegraphics[width=0.55\linewidth]{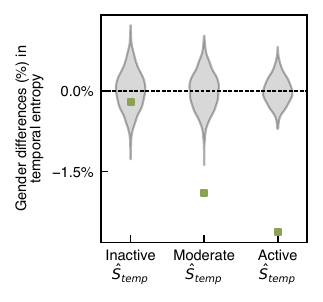}
    \caption{\textbf{Women exhibit higher temporal entropy across moderate and active groups, even after controlling for $N$ and $k$.}
        Relative differences in $\hat{S}^{temp}$ between women and men ($\frac{\text{male} - \text{female}}{(\text{male} + \text{female})/2}$), estimated via nearest-neighbour matching by activity and repertoire size (see Methods).
        Results are shown across activity groups (green squares).
        Grey violin plots show the reference distribution from 1,000 random shuffles of gender labels.}
    \label{fig:si-stemp}
\end{figure}

To assess gender differences in $\hat{S}^{temp}$, we applied nearest-neighbour matching to control for activity ($N$) and repertoire size ($k$) (see Methods).
Figure~\ref{fig:si-stemp} reports the relative gender differences ($\frac{\text{male} - \text{female}}{(\text{male} + \text{female})/2}$) across activity groups (inactive, moderate, active).

Women exhibit higher temporal entropy than men in both the moderate and active groups. Individuals in these groups have longer visit sequences and more unique states (locations) than those in the inactive group, indicating that the gender gap in temporal entropy widens as sequence length increases.

Therefore, beyond heterogeneity in visitation patterns, temporal ordering reveals gender differences not captured by visitation entropy alone.

\end{document}